\documentclass[adp,fleqn]{w-art}
\usepackage{times}
\usepackage{w-thm}
\usepackage[]{graphicx}
\usepackage{amssymb,amscd,amsmath,graphicx}
\chardef\bslash=`\\ 

\begin{document}
\DOIsuffix{theDOIsuffix}
\Volume{}
\Issue{}
\Copyrightissue{01}
\Month{}
\Year{2005}
\pagespan{1}{}
\Receiveddate{}
\Accepteddate{}



\title[Brownian motors]{Brownian motors}


\author[Peter H\"anggi]{Peter H\"anggi\footnote{Corresponding
     author \quad E-mail: {\sf peter.hanggi@physik.uni-augsburg.de}, Phone: +49\,821\,598\,3250,
     Fax: +49\,821\,598\,3222}\inst{1}}
\address[\inst{1}]{Universit\"at Augsburg, Institut f\"ur Physik, Universit\"atsstrasse 1, 86135 Augsburg, Germany}
\author[Fabio Marchesoni]{Fabio Marchesoni\inst{2}}
\address[\inst{2}]{Dipartimento di Fisica, Universit\`{a} di Camerino, 62032 Camerino, Italy}
\author[Franco Nori]{Franco Nori\inst{3,4}}
\address[\inst{3}]{Frontier Research System, The Institute of Physical and Chemical Research
(RIKEN), Wako-shi, Saitama, 351-0198, Japan}
\address[\inst{4}]{Center for Theoretical Physics, Department of Physics, University of Michigan,
Ann Arbor, MI 48109-1120, USA}


\begin{abstract}
In systems possessing a spatial or dynamical symmetry breaking
thermal Brownian motion combined with unbiased, non-equilibrium
noise gives rise to a channelling of chance that can be used to
exercise control over systems at the micro- and even on the
nano-scale. This theme is known as ``Brownian motor'' concept. The
constructive role of (the generally overdamped) Brownian motion is
exemplified for a noise-induced transport of particles within
various set-ups. We first present the working principles and
characteristics with a proof-of-principle device, a diffusive
temperature Brownian motor. Next, we consider very recent
applications based on the phenomenon of signal mixing. The latter is
particularly simple to implement experimentally in order to optimize
and selectively  control a rich variety of directed transport
behaviors. The subtleties and also the potential for Brownian motors
operating in the quantum regime are outlined and some
state-of-the-art applications, together with future roadways, are
presented.
\end{abstract}
\maketitle





\section{Introduction}
\label{Introduction}

In his {\it annus mirabilis} 1905,  Albert Einstein (March 14, 1879
- April 18, 1955) published four cornerstone papers that made him
immortal. Apart from his work on the photo-electric effect (for
which he obtained the Nobel prize in 1921), wherein he put forward
the photon hypothesis, and his two papers on special relativity, he
published his first paper on the molecular-kinetic description of
Brownian motion \cite{einstein1905a}. There, he states (freely
translated from the German) ``In this work we show, by use of the
kinetic theory of heat, that microscopic particles  which are
suspended in fluids undergo movements of such size that these can be
easily detected with a microscope. It is possible that these
movements to be investigated here are identical with  so-called
Brownian molecular motion; the information available to me on the
latter, however, is so imprecise  that I cannot make a judgement.''
In his follow up paper in 1906 \cite{einstein1906a}, which contains
the term ``Brownian motion'' in the title, he provides supplementary
technical arguments on his derivation and additionally  presents a
treatment of rotational Brownian motion. In this second paper he
also cites experimental work on Brownian motion by M. Gouy
\cite{Gouy1888} (but not Robert Brown). Einstein seemingly was
unaware of the earliest observations of Brownian motion under a
microscope: namely, the work of the Dutch physician Jan Ingen-Housz
\cite{ingenhousz1784a}, who detected, probably first, Brownian
motion of finely ground charcoal particles in a suspension at the
focal point of a microscope, and the detailed studies by the renown
 botanist Robert Brown \cite{brown1828a}. In clear contrast to
Robert Brown, who performed a series of experiments, Ingen-Housz
provided a quite incorrect physical explanation of his observations
by ascribing the effect to the evaporation of the suspension fluid.

The two founders of Brownian motion theory, Einstein and
Smoluchowski \cite{smoluchowski1906a}, as well as their
contemporaries, were also unaware of  related,
mathematical-statistical precursors of the phenomenon: Already in
1880, N. Thiele \cite{thiele1880a} proposed a model of Brownian
motion while studying time series. Another important development is
the work by the founder of modern Mathematical Finance, Louis
Bachelier \cite{bachelier1900a}, who attempted to model the market
noise of the Paris Bourse through a Gaussian process. Moreover, Lord
Rayleigh \cite{rayleigh1891a} also did study a discrete, heavy
random walker and performed a corresponding limiting procedure
towards a heat equation which is augmented by a drift term for the
statistical velocity.

These mathematical-statistical works already contain implicitly,
via the (Gaussian)-propagator solution  of the corresponding heat
or diffusion equation, the main result of Einstein: namely, his
pivotal analysis of the mean squared displacement of Brownian
motion. Einstein focused on what is nowadays characterized as {\it
overdamped Brownian motion}. He was driven by the quest for the
missing connection between macroscopic and molecular dimensions.
In doing so, his result exhibits truly remarkable features:
\begin{itemize}

\item The average distance traveled by the Brownian particle is
not ballistic. The latter only holds for transient, very short
times, typically of the order of $10^{-7}$ seconds, or smaller; an
estimate, which already Albert Einstein  provided in a subsequent short
note \cite{einstein1907a}. A benchmark result of Brownian motion
theory is that the average displacement (after the above mentioned
short transient)  is   proportional to $t^{1/2}$.
\end{itemize}

Thus, the velocity of a Brownian particle is not a useful measurable
quantity. Indeed,  earlier experimental attempts aimed at measuring
the velocity of Brownian particles, -- like those by Sigmund Exner
\cite{Sexner1867}, and many years later, the repeated, but far
better devised quantitative measurements by his son Felix Exner  in
1900 \cite{exner1900a}, -- yielded puzzling results, and
consequently were doomed to failure.

\begin {itemize}
\item Einstein also showed that the diffusion strength is related
to the Boltzmann constant (i.e., to the ratio of the ideal gas
constant $R$ and the Avogadro-Loschmidt  number $N_{A}$) and the
molecular dimension via the expression of Stokes' friction.
\end {itemize}

The last finding motivated Jean Perrin and collaborators
\cite{perrin1914a} to undertake detailed experiments on Brownian
motion, thereby accurately determining  the value for the
Avogadro-Loschmidt number.

The  relation described in this second feature also provides a first
link between dissipative forces and fluctuations. This {\it Einstein
relation} is a first example of the intimate relation between
thermal noise and dissipation that characterizes thermal
equilibrium: it is known under the label of the
fluctuation-dissipation theorem, put on firm ground only much later
\cite{callen1951a}.

It is just this overdamped Brownian noise which we attempt to
harvest with the concept of a Brownian motor
\cite{bartussek1995a,BM}. Put differently: can one extract  energy
from Brownian (quantum or classical) particles in asymmetric set-ups
in order to perform useful work against an external load? If true,
then it would be possible to rectify thermal Brownian motion so as
to separate, shuttle or pump  particles on a micro- or even
nano-scale. In view of the laws of thermodynamics, in particular the
second law,  the answer is obviously a firm {\em no}. If we could
indeed  succeed, then such a devilish device would constitute a
Maxwell demon perpetuum mobile of the second kind \cite{leff1990a}.
The only back door open is thus to go away from thermal equilibrium,
so that the constraints of thermodynamic laws no longer apply. This
leads us to study non-equilibrium statistical mechanics in
asymmetric systems. There, the symmetry is broken either (i) by the
system characteristics, such as an asymmetric periodic potential (or
substrate) which lacks reflection symmetry, called ratchet-like
potentials,  or (ii) the dynamics itself that may break the symmetry
in the time domain.

Clearly, noise-induced, directed transport in the presence of a
static bias is trivial.  It is also an everyday experience that
macroscopic, unbiased disturbances can cause directed motion. The
example of a self-winding wrist watch, or even windmills prove the
case. The challenge  becomes rather intricate when we consider
motion on the micro-scale. There, the subtle interplay of thermal
noise, nonlinearity, asymmetry, and {\it unbiased} driving of either
stochastic, or chaotic, or deterministic origin can indeed induce a
rectification of the noise, resulting in directed motion of Brownian
particles \cite{BM}. As a consequence, new roadways open up to
optimize and control transport on the micro- and/or nano-scale. This
includes novel applications in physics, nano-chemistry, materials
science, nano-electronics and, prominently, also for directed
transport in biological systems such as in molecular motors
\cite{julicher1997a}. In the next section, the concept of a Brownian
motor will be illustrated with a diffusive Brownian motor.

\section{Archetype model of a Brownian motor}
\label{archetype} 
\begin{figure}[htb]
\includegraphics[width=.60\textwidth]{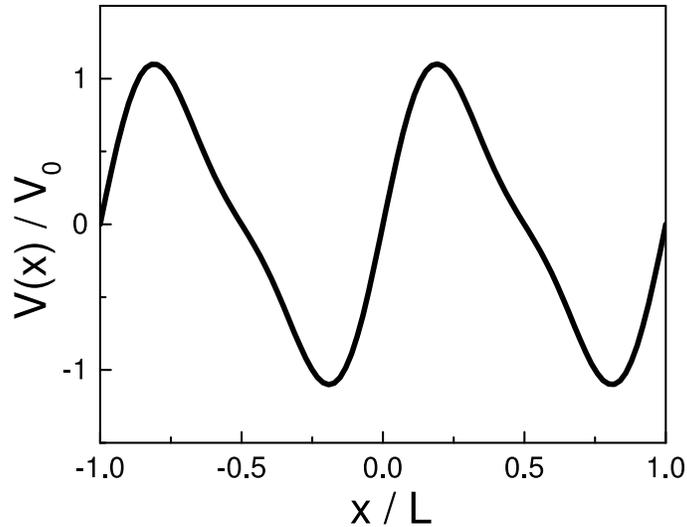}
\centering \hfil \caption{Typical example of a ratchet-potential
$V(x)$. It is periodic in the spatial coordinate with period $L$
and exhibits a broken spatial symmetry. Plotted here is the
example from (\ref{2.3}), in dimensionless units.} \label{fig:1}
\end{figure}

In order to elucidate the {\it modus operandi} of
a Brownian motor, we consider  a Brownian particle with mass $m$ and
friction coefficient $\eta$ in one dimension with coordinate $x(t)$,
being driven by an external static force $F$ and thermal noise. The
corresponding stochastic dynamics thus reads:
\begin{equation}
m\ddot x =   - V'(x) - \eta\,\dot x + F + \xi (t)\ , \label{2.1}
\end{equation}
where $V(x)$ is a periodic potential with period $L$,
\begin{equation}
V(x+L)=V(x)\ ,
\label{2.2}
\end{equation}
which exhibits broken spatial
symmetry (a so-called ratchet potential).
A typical example is
\begin{equation}
V(x) = V_0 \, [\sin (2\pi x/L) + 0.25\, \sin(4 \pi x/L)] \ ,
\label{2.3}
\end{equation}
which is depicted in Fig. \ref{fig:1}. \vglue 0.5cm
Thermal fluctuations are modelled by a Gaussian white noise of
vanishing mean, $\langle\xi(t)\rangle = 0$, satisfying Einstein's
{\em fluctuation-dissipation} relation, i.e.
\begin{eqnarray}
\langle\xi(t)\xi(s)\rangle & = &
2\,\eta\, k_{B}  T \, \delta(t-s) \ ,
\label{2.4}
\end{eqnarray}
where $k_B $ is the Boltzmann constant and $T$ denotes the
equilibrium temperature.

In extremely small systems, particle fluctuations are often
described to a good approximation by the {\it overdamped} limit of
Eq. (\ref{2.1}), i.e., by the Langevin equation
\begin{equation}
\eta\,\dot x =   - V'(x) + F + \xi (t)\ , \label{2.1bis}
\end{equation}
where the inertia term $m\ddot x$ has been neglected altogether
[as implicit in Einstein's work].

In the absence of an external bias, i.e. $F=0$, the second law of
thermodynamics implies that the thermal equilibrium stochastic
dynamics cannot support a stationary current, i.e., $\langle \dot
x(t) \rangle = 0$. This  can be readily proven \cite{BM} upon
solving the corresponding Fokker-Planck equation in the space of
periodic probability functions, with the stationary probability
being of the Boltzmann form.

This pivotal result no longer holds, however, when we complement
our archetype model by a non-equilibrium, {\it unbiased} (i.e.
zero mean) disturbance. An instructive way consists in applying a
temporally varying temperature $ T \rightarrow T(t)$, with $T(t)$
being a periodic function in time \cite{reimann1996a}. This means
that the Einstein relation is modified to read
\begin{equation}
\langle\xi(t)\xi(s)\rangle = 2\,\eta\, k_{B}  T(t) \,
\delta(t-s)\, , \label{2.5}
\end{equation}
with the temperature obeying $ T(t) = T (t + {\cal T})$, where
${\cal T}$ denotes the period of the temperature modulation. Most
importantly, such an explicit time dependence moves the system {\it out
of thermal equilibrium}. In particular, the system dynamics is no
longer time-homogeneous; it thus breaks also the detailed balance
symmetry \cite{hanggi1982a}. Note that this latter symmetry must
always be obeyed in thermal equilibrium.
A typical periodic temperature modulation is:
\begin{equation}
T(t) =  {T} \, (1+A\,{\rm sgn}[\sin (2\pi t/{\cal T})])
\label{2.6}
\end{equation}
where ${\rm sgn}[x]$ denotes the signum function, and $|A|<1$. This
variation of the temperature $T$ in (\ref{2.6}) causes jumps of
$T(t)$ between $T_{\rm hot}= {T} \, (1+A)$ and $T_{\rm cold}= {T} \,
(1-A)$ at every half-period ${\cal T}/2$. Due to these cyclic
changes of the temperature, the system approaches a periodic
long-time stationary state which, in general, can  be investigated
only numerically in terms of Floquet theory \cite{reimann1996a}.
\vglue .5cm
\begin{figure}[htb]
\includegraphics[width=.85\textwidth]{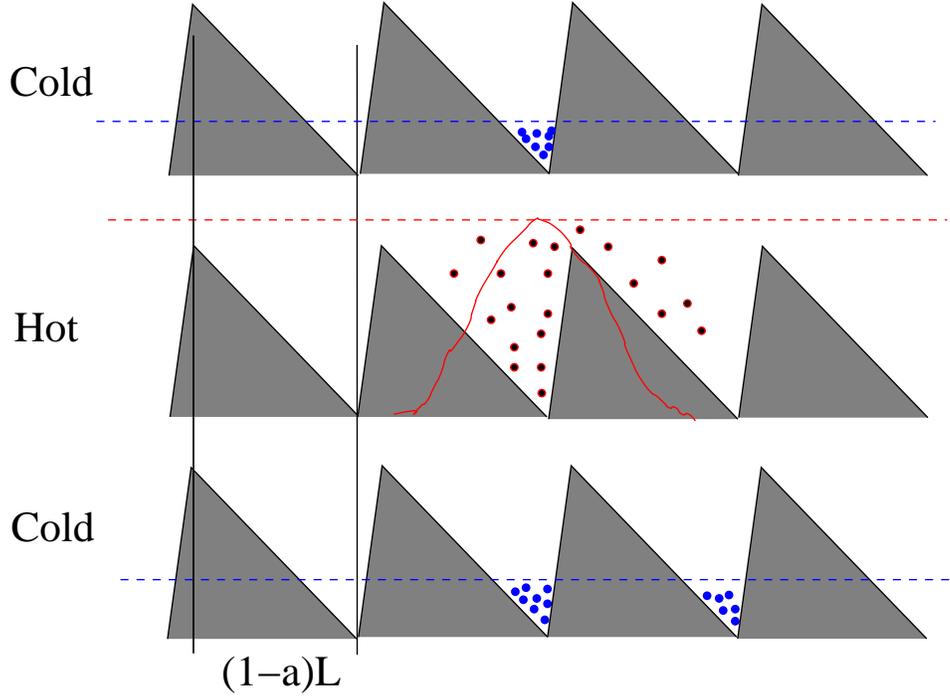}
\centering \hfil \caption{Working principle of a Brownian motor
driven by temperature oscillations \cite{reimann1996a}: Consider
suspended noninteracting particles in a viscous medium moving along
an asymmetric saw-tooth potential of period $L$ and height $\Delta
V$ which are subjected to a temperature that changes in time between
the values ``hot'' and `` cold'' , $T \in [T_{\rm cold},T_{\rm
hot}]$. For simplicity, let $\Delta V/(k_B T_{\rm hot}) \ll 1$, and
$\Delta V/(k_B T_{\rm cold}) \gg 1$. The dashed line in the middle
panel indicates the level $2k_B T$ below which circa 95 \% of the
particles are found at any given time. Initially, when the
temperature is cold, the particles are pinned at a potential
minimum. Then, when the temperature is switched to ``hot'', the
particles effectively do not feel the potential and begin to
diffuse. In the middle illustration the thin red line  indicates a
Gaussian-like shape for the corresponding  particle distribution.
When the temperature is low again, any particles that have diffused
the short distance $L_1=aL$, with $a < 1/2$, to the right are caught
in the well to the right; likewise, any particles that have diffused
the long distance $L_2=(1-a)L$ towards the left are caught in the
well to the left, and the rest are pinned again in the original well
from which they started out. Because the chance for a particle to
diffuse over the short distance $L_1$ during the time when the
temperature is high is much larger than the chance to diffuse over
the long distance $L_2$, a net motion to the right is induced by
such cyclic temperature fluctuations.} \label{fig:2}
\end{figure}

In the case of a static tilted Brownian motor with a fixed
temperature $T$, we  immediately see that for a given force, say
$F<0$, the particle will on average move ``downhill'', i.e.
$\langle\dot x\rangle < 0$. This fact holds true for {\em any
fixed}, non-zero value of the temperature $T$. Returning to the
temperature ratchet with $T$ being  subjected to periodic, temporal
variations, one  should expect that the particles still move
``downhill'' on the average. The numerically calculated corresponding ``load
curve'' (see Fig. 2 in \cite{BMRH})  demonstrates, however, that the
 opposite is true within an entire interval of {\it negative} bias
values $F$: Surprisingly indeed, {\em the particles are climbing
``uphill'' on the average, thereby performing work against the
load force $F$}. This upward directed motion is apparently
triggered by no other source than the thermal fluctuations
$\xi(t)$. This key finding is just what is commonly referred to as
the {\it Brownian motor effect} \cite{BM,BMRH}.

Because the average particle current $\langle\dot x\rangle$
depends continuously on the load force $F$,
it is sufficient for a qualitative analysis  to consider the
case $F=0$: {\em the occurrence of the
Brownian motor or ratchet effect  is then tantamount to a finite
current}
\begin{equation}
\langle\dot x\rangle\not = 0 \ \ \ \mbox{for}\ \ \  F=0 \ ,
\label{2.7}
\end{equation}
i.e., the unbiased Brownian motor
implements a {\em  directed motion of particles}.

\subsection{Working principle of a Brownian motor}

In order to understand the basic physical mechanism behind the
ratchet effect at $F=0$, we focus on very strong, i.e. $ |A|
\lesssim 1$ but adiabatically slow, periodic two-state  temperature
modulations from (\ref{2.6}). During a first time interval, say $t
\in [{\cal T}/2,{\cal T}]$, the thermal energy $k_B  T(t)$ is kept
at a constant value $k_B  {T} \, (1-A)$ much {\it smaller} than the
potential barrier $\Delta V$ between two neighboring local minima of
$V(x)$. Thus, all particles will have accumulated in a close
vicinity of the potential minima at the end of this time interval,
as sketched in the top panel of Fig. \ref{fig:2}. Then, the thermal
energy jumps to a value $ k_B  {T} \, (1+A)$ much {\it larger} than
$\Delta V$ and remains stable during another half period, say $t \in
[{\cal T}, 3 {\cal T}/2]$. Because the particles then barely feel
the potential profile in comparison to the intense noise level, the
particles spread out subject to free thermal diffusion -- see Fig.
\ref{fig:2}, middle panel. Finally, $T(t)$ jumps back to its
original ``cool'' value ${T} \, (1-A)$ and the particles slide
downhill towards the closest local minima of $V(x)$. Due to the lack
of reflection symmetry of the function $V(x)$, the original
population of one given potential well is thus re-distributed
asymmetrically, yielding a net average displacement after one
temporal period ${\cal T}$.

When the temperature is varied very slowly during a cycle (and
restricting the  discussion to the case that the potential $V(x)$
has only one minimum and maximum per period $L$, like in Fig.
\ref{fig:2}) it is quite obvious that if the local minimum is closer
to its adjacent maximum located to the right, a positive particle
current $\langle\dot x\rangle >0$ will arise. Put differently, upon
inspection  of Fig. \ref{fig:2}, it is intuitively clear that during
the cool-down cycle the particles must diffuse a long distance to
the left, but only a short distance to the right. This in turn
induces a net transport against the steeper potential slope towards
the right. All these predictions rely on our assumptions that ${T}
\, (1\mp A)$  are much smaller/larger than $\Delta V$, and that the
time-period $\cal T$ is sufficiently large.

The Brownian motor effect (\ref{2.7}) occurs for very general
temperature modulations $T(t)$, as well. For the same reason, the
ratchet effect is also robust with respect to modifications of the
potential shape \cite{reimann1996a} and is recovered even for
random instead of deterministic modulations of $T(t)$
\cite{temperatureBM}, with a modified dynamics on a discrete state
space \cite{sokolov1997a}, and in the  presence of  finite inertia
\cite{bao2000a}.

The directed particle current is clearly bound to vanish in the so
termed adiabatic limit (i.e. for  asymptotically, very slow
temperature modulations),  when thermal equilibrium is approached. A
similar conclusion holds true for asymptotically fast temperature
modulations. By use of a correspondent,  perturbative Floquet
analysis one finds the noteworthy result that the current decays to
zero in both asymptotic regimes remarkably fast, namely like ${\cal
T}^{-2}$ in the slow modulation limit, and ${\cal T} ^2$, in the
fast modulation limit, respectively \cite{reimann1996a}.

Moreover, for non-adiabatic temperature variations, the Brownian
motion in a diffusive Brownian motor moving on a tailored ratchet
profile is generally not rectified in its ``natural'' direction, but
rather in the opposite direction \cite{reimann1996a}. This in turn
implies a time-scale induced (non-adiabatic) current-reversal: It is
this very  feature that is required for an efficient separation of
particles of different size, or other transport qualifiers such as
friction, mass, etc..

\subsection{Features of a Brownian motor}

We cannot emphasize enough that the ratchet effect, as exemplified
in the temperature Brownian motor model shown in Fig. \ref{fig:2},
is {\em not} in contradiction with the second law of thermodynamics:
the temperature changes in (\ref{2.6}) are caused by two heat
environments at two different temperatures with which the Brownian
motor system is in continuous contact. From this viewpoint, this
archetype Brownian motor is nothing else than an extremely simple,
small heat engine. The fact that such a device can produce work is
therefore not a miracle --- but it is still very intriguing. The
following characteristics are a hallmark  of Brownian motors.

\subsubsection{Loose-coupling mechanism}

Consider the ``relevant state variables'' $x(t)$ and $T(t)$ of our
temperature Brownian motor. In the case of an ordinary heat
engine, these two state variables would always cycle through one
and the same periodic sequence of events (``working strokes'').
Put differently, the evolutions of the state variables $x(t)$ and
$T(t)$ would be tightly coupled  and almost synchronized.

In clear contrast to this familiar scenario, {\em the relevant state
variables of a genuine Brownian motor are loosely coupled}: Some
degree of interaction is required for the functioning of the
Brownian motor, but while $T(t)$ completes one cycle, $x(t)$ may
evolve in a very different way. The spatial coordinate $x(t)$ is
certainly not {\it slaved} by the unbiased modulation of the
temperature $T(t)$.

This {\em loose coupling between state variables} is a salient
feature of a Brownian motor device and distinguishes the Brownian
motor concept from micron-sized, but otherwise quite conventional
thermo-mechanical or even purely mechanical engines. In
particular, indispensable ingredients of {\it any} genuine
Brownian motor are: (i) the presence of some amount of (not
necessarily thermal) noise; (ii) some sort of symmetry-breaking
supplemented by temporal periodicity (possibly via an unbiased,
non-equilibrium forcing), if a cyclically operating device is
involved. It is thus  not appropriate to advertise every such
small ratchet device under the trendy label of ``Brownian motor''.
This holds true especially if the governing transport principle is
deterministic, like in mechanical ratchet devices of macro- or
mesoscopic size, such as a ratchet wrench, interlocked mechanical
gears, or Leonardo's ``cochlea'' \cite{leonardo1} and other
``screw-like'' pumping and propulsion devices. By the way, it is
suggestive to notice how Leonardo sketched a ratchet-like
machinery just to prove the impossibility of the {\it perpetuum
mobile}  \cite{leonardo2}.

\subsubsection{Dominant role of noise}
Yet another distinguishing feature of a Brownian motor is that
noise (no matter what its source, i.e. stochastic, or chaotic, or
thermal) plays a non-negligible, or even a dominant role. In
particular, it is the intricate interplay among nonlinearity,
noise-activated escape dynamics and non-equilibrium driving which
implies, that, generally, {\em not even the direction of
transport} is {\it a priori} predictable. See also in Sects. 4 and
5 below.

\subsubsection{Necessary ingredients and variations of the Brownian motor scheme}
The necessary condition for the Brownian motor effect is to
operate away from thermal equilibrium, namely, in a state with no
detailed balance. This was achieved above through the cyclic
variation of the temperature (\ref{2.6}); but there clearly exists
a great variety of other forms of non-equilibrium perturbations
\cite{others}. The following guiding prescriptions should be
observed when designing a more general {\em Brownian motor}:

\begin{itemize}
\item Spatial and/or temporal periodicity critically affect
rectification.
\item All acting forces and gradients must vanish
after averaging over space, time, and statistical ensembles.
\item Random forces (of thermal, non-thermal, or even deterministic
origin) assume a prominent role.
\item Detailed balance symmetry
must be broken by moving the system away from thermal equilibrium.
\item  A symmetry-breaking must apply.
\end{itemize}

There exist several possibilities to induce  symmetry-breaking.
First, the spatial inversion symmetry of the periodic system itself
may be broken {\em intrinsically}; that is, already in the absence
of  non-equilibrium perturbations. This is the most common situation
and typically involves a type of periodic asymmetric (so-called
ratchet) potential. A second option consists in the use of a
 deterministic, unbiased skew forcing $f(t)$. For example, these
may be {\em stochastic} fluctuations $f(t)$ possessing
non-vanishing, higher order odd multi-time moments ---
notwithstanding the requirement that they must be unbiased, i.e.
the first moment vanishes \cite{hanggi1996a}. Such an asymmetry
can also be created by unbiased {\em periodic} non-equilibrium
perturbations $f(t)$. Both variants in turn induce a spatial
asymmetry of the dynamics. Yet a third possibility arises via a
collective effect in coupled, perfectly symmetric non-equilibrium
systems, namely in the form of {\em spontaneous symmetry breaking}
\cite{cooperative,cooperative2,cooperative3}. Note that in the
latter two cases we speak of a Brownian motor dynamics even though
a ratchet-potential is not necessarily involved. An instructive
demo java applet of a Brownian motor can be found  on the web
\cite{Elmer}.

In the next section we shall illustrate this concept for the case of
a temporal, {\it dynamical symmetry breaking}. This approach is
readily implemented experimentally, and therefore does carry a great
potential for novel applications and devices.

\section{Brownian motors and dynamical symmetry breaking} \label{Sec1}

What we call a ratchet mechanism is to some extent a matter of
taste. Our {\it bona fide} ratchet prescriptions in Sec. 2.2 apply
indeed to a number of diverse microscopic rectification mechanisms
that have been known for a long time, well before the notion of
Brownian motor became popular. Such mechanisms do involve
ingredients like spatial periodicity, random forces,
non-equilibrium (detailed balance breaking) and zero-mean external
biases (both in space and time). However, in this category of
processes the reflection asymmetry of the substrate plays no
essential role, although its presence may add certain sofar
unnoticed similarities with the ratchet phenomenology. A common
feature of all these non-ratchet, or possibly ratchet-related,
rectification mechanisms is some degree of temporal
synchronization between input signal(s) and/or spatial modulation
of the substrate, leading to a dynamical symmetry breaking.

\subsection{Harmonic mixing} \label{SS1}

A charged particle spatially confined by a nonlinear force is
capable of mixing two alternating input electric fields of angular
frequencies $\Omega_1$ and $\Omega_2$, its response containing all
possible higher harmonics of $\Omega_1$, $\Omega_2$ and their sum
and difference frequencies. For commensurate input frequencies,
i.e., $m\Omega_1=n\Omega_2$, there appears a rectified output
component, too \cite{oldHM}: such a dc harmonic mixing (HM) signal
is an $(n+m)$-th order effect in the dynamical parameters of the
system \cite{marchesoni1986a,goychuk1998}.

Let us consider the stochastic dynamics of an overdamped particle with coordinate $x(t)$,
\begin{equation}
\label{HM1} \dot x=-V'(x) +F(t) +\xi(t),
\end{equation}
moving on the one dimensional substrate potential $V(x)=q(1-\cos
x)$, subjected to an external zero-mean Gaussian noise $\xi(t)$ with
the auto-correaltion function $\langle \xi(t)\xi(0)\rangle =
2D\delta(t)$, and a periodic two-frequency drive force
\begin{equation}
\label{HM2} F(t)=A_1\cos(\Omega_1 t+\phi_1) +A_2\cos(\Omega_2
t+\phi_2),
\end{equation}
with $\Omega_1$ and $\Omega_2$ integer-valued multiples of the
fundamental frequency $\Omega_0$, i.e., $\Omega_1=n\Omega_0$ and
$\Omega_2=m\Omega_0$. For $D=k_BT$ the Langevin equation
(\ref{HM1}) is the zero-mass limit of Eq. (\ref{2.1}) with
$\eta=1$; in the case of bistable potentials it describes a
well-known synchronization phenomenon known as Stochastic
Resonance, both in physics \cite{SRrmp} and biology \cite{SRcpc}.
A standard perturbation expansion leads to a general expression
\cite{marchesoni1986a} for the non-vanishing dc component $j_0
\equiv \langle \dot x \rangle$ of the particle velocity. In the
regime of low temperature, $D\ll \Delta V=2q$, the particle net
current can be approximated to
\begin{equation}
\label{HM3} \frac{j_0}{D}=-\frac{\pi^2}{2^{m+n}}
\left(\frac{A_1}{D}\right)^m
\left(\frac{A_2}{D}\right)^n\cos(n\phi_2-m\phi_1)
\end{equation}
for $\Omega_0^2\ll q$ (low frequency limit), and to
\begin{equation}
\label{HM4} \frac{j_0}{\Omega_0}=-\frac{1}{2^{m+n}}
\left(\frac{q}{D}\right)^2 \left(\frac{A_1}{\Omega_1}\right)^m
\left(\frac{A_2}{\Omega_2}\right)^n\cos(n\phi_2-m\phi_1)
\end{equation}
for $\Omega_0^2\gg q$ (high frequency limit). The sign of $j_0$ is
controlled by the input phases $\phi_1$, $\phi_2$, while an
average over $\phi_1$ or $\phi_2$ would eliminate the
rectification effect completely. The two sinusoidal components of
$F(t)$ get coupled through the anharmonic terms of the substrate
potential $V(x)$ \cite{breymayer1981}; the dependence of $j_0$ on
$\Delta_{m,n}=n\phi_2-m\phi_1$ characterizes HM indeed as a
synchronization effect.

In the derivation of Eqs. (\ref{HM3}) and (\ref{HM4}) no
assumption was made regarding the reflection symmetry of $V(x)$;
actually, HM rectification may occur on symmetric substrates, too.
However, a simple perturbation argument \cite{landau} leads to the
conclusion that a symmetric device cannot mix low-frequency
rectangular waveforms, namely no HM is expected for
\begin{equation}
\label{HM5}F(t)=A_1\hbox{sgn}[\cos(\Omega_1 t+\phi_1)]
+A_2\hbox{sgn}[\cos(\Omega_2 t+\phi_2)],
\end{equation}
with $A_1,A_2 \geq 0$ and $\hbox{sgn}[\dots]$
denoting the sign of $[\dots]$. However, an {\it asymmetric}
device can!

In order to illustrate the properties of {\it asymmetric} HM
\cite{savelev2004a} let us consider the driven dynamics
(\ref{HM1}) in the piecewise linear potential $V(x)=x\Delta V/L_2$
for $-L_2 <x <0$ and $V(x)=x\Delta V/L_1$ for $0<x <L_1$, with
$L_1+L_2=L$ and, say, $L_2<L_1$, i.e. opposite polarity with
respect to the potential in Fig. \ref{fig:2}; the external drive
$F(t)$, in Eq.~(\ref{HM5}), is assumed to vary slowly in time.

The advantage of imposing the adiabatic limit $\Omega_1$, $\Omega_2
\rightarrow 0$, is that the output $j(\Omega_1,\Omega_2, A_1, A_2)$
of a such doubly-rocked ratchet is expressible analytically in terms
of the current $j_{R}(A)$ of the standard one-frequency rocked
ratchet \cite{bartussek1994a}, obtained by setting $A_1=A$ and
$A_2=0$ in Eq. (\ref{HM5}). Note that here $j_{R}(A)$ is a symmetric
function of $A$, and in the adiabatic approximation $j_{R}(A)=
j_{R}(-A)= A[\mu(A)-\mu(-A)]/2$, where $\mu(A)$ is the mobility of
an overdamped particle running down the tilted potential $V(x)-Ax$.

The overall ratchet current $j(\Omega_1, \Omega_2, A_1, A_2)$
results from the superposition of the two standard one-frequency
currents $j_{R}(A_1+A_2)$ and $j_{R}(A_1-A_2)$ for drive ac
amplitudes $A_1+A_2$ and $A_1-A_2$, respectively
\cite{savelev2004a}. In particular, for any positive integers $m$,
$n$ with $m>n$,
\begin{eqnarray}
\label{HM6} j(\Omega_1, \Omega_2=\Omega_1\frac{2m-1}{2n-1}, A_1,
A_2)& = & j_{\rm avg}(A_1, A_2) \nonumber  \\ & &
-{{(-1)^{m+n}}\over{(2m-1)(2n-1)}}~\Delta j(A_1,A_2)
~p(\Delta_{n,m}),
\end{eqnarray}
where
\begin{eqnarray}
\label{HM7} & & j_{\rm avg}(A_1, A_2) = \frac{1}{2}
[j_{R}(A_1-A_2)+j_{R}(A_1+A_2)] \;, \nonumber \\ & &  ~~~~ \Delta
j(A_1,A_2)= \frac{1}{2} [j_{R}(A_1-A_2)-j_{R}(A_1+A_2)],
\end{eqnarray}
and $p(\Delta_{n,m})={|\pi-\Delta_{n,m}|}/{\pi}-0.5$ is a
modulation factor with $\Delta_{n,m}= (2n-1)\phi_2-(2m-1)\phi_1$,
mod($2\pi$).

\begin{figure}[htb]
\centering
\includegraphics[width=1.0\textwidth]{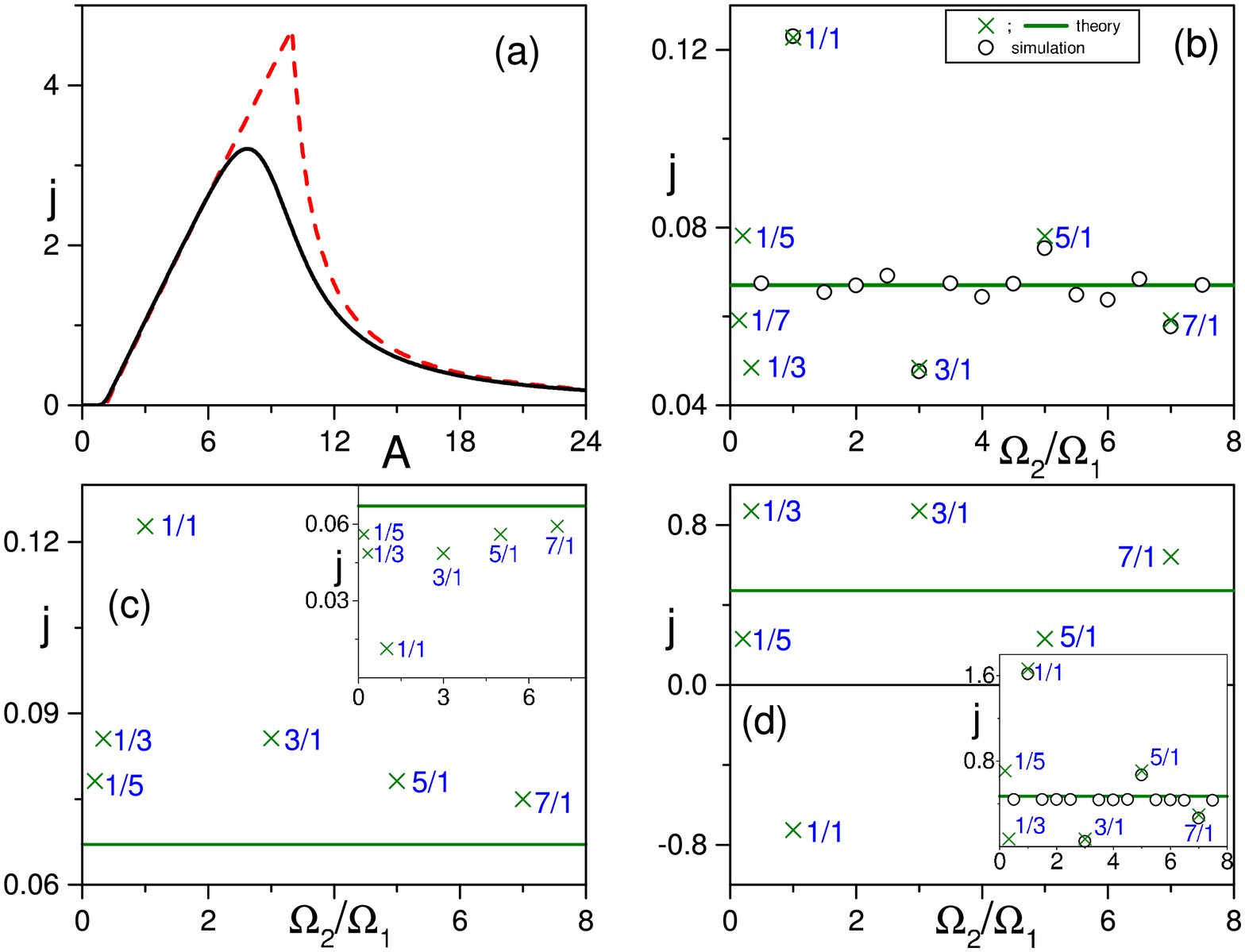}
\caption{Rectified current driven by two rectangular waveforms with
fixed amplitudes: (a) one-frequency rocked ratchet; (b), Harmonic
mixing case of Eqs. (\ref{HM1}), (\ref{HM2}); (d) gating mechanism
(\ref{G1}). Panel (a): Response curve $j_R(A)$ of the potential
$V(x)$ driven by a low-frequency rectangular force with amplitude
$A$ at zero temperature $D = 0$ (dashed curve), and  low temperature
$D/\Delta V= 0.05$ (solid curve). Panel (b): Numerical simulations
for a doubly rocked ratchet with $\phi_1=\phi_2=\pi$ and
$\Omega_1=0.01$ (open circles) and in adiabatic approximation (green
line and green crosses). The baseline $j_{\rm avg}$, Eq.
(\ref{HM7}), is indicated by the green line; the spikes at some
selected integer-valued odd harmonics are marked with green crosses
($\times$); Panel (c): Adiabatic approximation for
$\phi_1=\phi_2=3\pi/2$ (main panel) and $\phi_1=3\pi/2$,
$\phi_2=\pi/2$ (inset). In both cases $A_1=3$, $A_2=2$, $D=0.6$;
Panel (d): Numerical simulations for a rocked-pulsated ratchet in
the adiabatic regime with $A_1=4$, $A_2=0.5$ and $\Omega_1=0.01$;
noise level: $D=0.4$. Main panel: $\phi_1=\phi_2=\pi$ (adiabatic
approximation); inset: simulation (open circles) versus the fully
adiabatic approximation ($\times$) for $\phi_1=\pi$ and $\phi_2=0$.
$V(x)$ parameters are: $L_1=0.9$, $L=1$, $\Delta V=1$ in (a)-(c) and
$\Delta V=2$ in (d).}\label{FigHM1}
\end{figure}
The most significant properties of the rectification current
(\ref{HM6}) and (\ref{HM7}) are elucidated in Fig.
\ref{FigHM1}(a), (b), where results from numerical simulation are
displayed for a comparison \cite{savelev2004a}: (1) The doubly
rocked ratchet current in the adiabatic limit) is insensitive to
$\Omega_1$, $\Omega_2$ for $\Omega_2 \neq \Omega_1
=(2m-1)/(2n-1)$;
 its intensity coincides with the
``baseline" value $j_{\rm avg}(A_1, A_2)$ of Eq.~(\ref{HM7}); spikes
correspond to odd fractional harmonics; their amplitude $\Delta
j(A_1,A_2)/(2m-1)(2n-1)$ is suppressed at higher harmonics, i.e.,
for larger $m,n$. (2) The sign of the spike factor $\Delta
j(A_1,A_2)$ is sensitive to the signal amplitudes $A_1$, $A_2$. For
instance, if we choose $A_1$, $A_2$ so that $A_1+A_2$ and
$|A_1-A_2|$ fall onto the rising (decaying) branch of $j(A)$ in Fig.
\ref{FigHM1}a, then $\Delta j(A_1,A_2)$ is negative (positive). (3)
The current spikes at $\Omega_2/\Omega_1=(2m-1)/(2n-1)$ depend on
the initial value of $\phi_1$, and for a fixed $\phi_1$, their
amplitude oscillates with $\phi_2$ proportional to the modulation
factor $p(\Delta_{n,m})$. We remark that the overall sign of our
doubly rocked ratchet current is always determined by the polarity
of $V(x)$ (positive for $L_2<L_1$), as $|\Delta j(A_1,A_2)| <
|j_{\rm avg}(A_1,A_2)|$ for any choice of $A_1$, $A_2$. However, in
the {\it partially adiabatic} regime, where only one frequency tends
to zero, multiple current inversions are also possible
\cite{savelev2004a}.

\subsection{Gating mechanism} \label{SS2}

A periodically-driven Brownian motion can also be rectified by
modulating the amplitude of the substrate potential $V(x)$. Let us
consider for instance the overdamped dynamics described by the
Langevin equation
\begin{equation}
\label{G1} \dot x=-V'(x)[1+F_2(t)] +F_1(t) +\xi(t).
\end{equation}
To avoid interference with possible HM effects we follow the
prescription of Sec. \ref{SS1}, namely we take $F_i(t)=A_i
\hbox{sgn}[\cos(\Omega_it+\phi_i)]$, with $i=1,2$ and $A_i\geq 0$.
Mixing of the additive $F_1(t)$ and the multiplicative signal
$F_2(t)$ provides a control mechanism of potential interest in
device design. Without loss of generality, our analysis can be
conveniently restricted to the piecewise linear potential $V(x)$
also used in the previous subsection.

In the adiabatic limit, the ac driven Brownian particle $x(t)$ can
be depicted as moving back and forth over a time modulated
potential $V(x,t)=V(x)[1+F_2(t)]$ that switches between two
alternating configurations $V_{\pm}(x)=V(x)(1 \pm A_2)$. Both
substrate profiles $V_{\pm}(x)$ are capable of rectifying the
additive driving signal $F_1(t)$ with characteristic functions
${j}_{\pm}(A_1)$, respectively; the net currents ${j}_{\pm}(A_1)$
are closely related to the curve $j_{R}(A)$ plotted in Fig.
\ref{FigHM1}(a), namely \cite{savelev2004a}
\begin{equation}
{j}_{\pm}(A_1)= (1\pm A_2)\, j_{R}\left[\frac{A_1}{1\pm
A_2}\right]\;,
\end{equation}
with $D\rightarrow D/(1\pm A_2)$. It follows that the total net
current can be cast in the form (\ref{HM6}) with
\begin{equation}
\label{G2} j_{\rm avg}(A_1, A_2)= (1/2)[{j}_-(A_1)+ {j}_+(A_1)]
\end{equation}
and
\begin{equation}
\label{G3} \Delta j(A_1,A_2)= (1/2)[{v}_-(A_1)- {v}_+(A_1)],
\end{equation}
where $v_{\pm}(A_1)=A_1[\mu_{\pm}(A_1)+\mu_{\pm}(-A_1)]/2$. We
recall that in our notation $\mu_{\pm}(A)$ is the static nonlinear
mobility of the tilted potentials $V_{\pm}(x)-Ax$.

It is apparent that $|\Delta j(A_1,A_2)|$ may grow larger than
$|j_{\rm avg}(A_1,A_2)|$ and, therefore, a current reversal may take
place for appropriate values of the model parameters, as shown by
the simulation results in Fig.~\ref{FigHM1}(d). In fact, already a
relatively small modulation of the ratchet potential amplitude at
low temperatures can  reverse the polarity of the simply rocked
ratchet $V(x)$. Let us consider the simplest possible case,
$\Omega_1=\Omega_2$ and $\phi_1=\phi_2$: As the ac drive is oriented
along  the ``easy" direction of $V(x)$, namely to the right, the
barrier height $V(x,t)$ is set at its maximum value $\Delta
V(1+A_2)$; at low temperatures the Brownian particle cannot overcome
this barrier height within a half ac-drive period $\pi/\Omega_1$. In
the subsequent half period the driving signal $F_1(t)$ changes sign,
thus pointing against the steeper side of the $V(x,t)$ wells, while
the barrier height drops to its minimum value $\Delta V(1-A_2)$:
Depending on the value of $\Delta V/D$, the particle has a better
chance to escape a potential well to the left than to the right,
thus making a current reversal possible. Of course, the net current
may be controlled via the modulation parameters $A_2$ and $\phi_2$,
too (see inset of Fig.~\ref{FigHM1}d).

Note that Eq. (\ref{HM6}) is symmetric under $m \leftrightarrow n$
exchange. This implies that, as long as the adiabatic
approximation is tenable, each spectral spike $(m,n)$ of the net
current is mirrored by a spike $(n,m)$ of equal strength (see Fig.
\ref{FigHM1}). This is not true, e.g., in the partially adiabatic
regime, where the dynamics depends critically on which ratio
$\Omega_1/\Omega_2$ or $\Omega_2/\Omega_1$ tends to zero
\cite{savelev2004a}.

The rectification effect introduced in this subsection rests upon
a sort of {\it dynamical symmetry breaking} mechanism, or {\it
synchronized gating}, which requires no particular substrate
symmetry. In the case of a symmetric piecewise linear potential,
$L_1=L_2$, the baseline current $j_{\rm avg}(A_1, A_2)$ clearly
vanishes, while the current spikes due to gating remain.

Asymmetry-induced and nonlinearity-induced mixing are barely
separable in the case of {\it sinusoidal} input signals. This case
is analytically less tractable and shows significant differences
with respect to the square-wave rectification investigated so far.
Spikes in the output current spectrum occur for any rational value
of $\Omega_2/\Omega_1=m/n$, including {\it even} fractional
harmonics, i.e. $\Omega_2/\Omega_1= 2m/(2n-1)$, or
$\Omega_2/\Omega_1=(2m-1)/2n$, respectively, but they are no longer
symmetric under the exchange of $m \leftrightarrow n$. This is so
because HM cannot be separated from asymmetry-induced mixing. It has
been noticed that a binary mixture of particles \cite{savelev2004b},
diffusing through a quasi-one dimensional channel, provides a
convenient study case (both numerical and experimental) to contrast
nonlinearity versus asymmetry induced signal mixing.

\subsection{Stokes' drift} \label{SS3}

Particles suspended in a viscous medium traversed by a longitudinal
wave travelling in the $x$-direction, are dragged along according to
a deterministic mechanism known as Stokes' drift \cite{stokes1847}.
As a matter of fact, the particles spend slightly more time in
regions where the force acts parallel to the direction of
propagation than in regions where it acts in the opposite direction.
Consider \cite{vandenBroeck1999a} a symmetric square wave
$f(kx-\Omega t)$ with wavelength $\lambda=2\pi/k$ and temporal
period $T_{\Omega}=2\pi/\Omega$, capable of entraining the particles
with velocity $\pm bv$ [with $v=\Omega/k$, $0<b<1$, and the signs
$\pm$ denoting the orientation of the force]. During one cycle, the
particle velocity is positive for a longer time interval,
$\lambda/2(1-b)v$, than it is negative, $\lambda/2(1+b)v$; hence,
the average drift velocity $v_S= b^2v$. The unknown factor $b$
depends on the speed of travelling wave $f(kx-\Omega t)$ and the
temperature of the propagation medium.

\begin{figure}[htb]
\centering
\includegraphics[width=.65\textwidth,clip=]{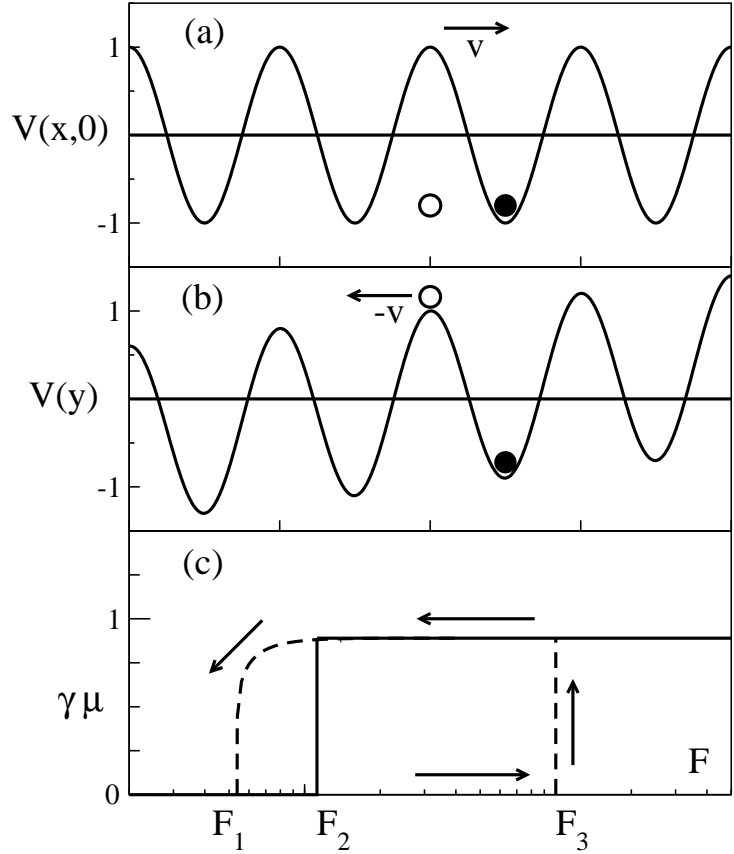}
\caption{(a) Snapshot of the travelling potential $V(x,t)$ of Eq.
(\ref{S2}) at $t=0$. (b) Tilted washboard potential $V(y)$ of Eq.
(\ref{S3}). A Brownian surfer is represented by a filled circle, a
Brownian swimmer by an empty circle. (c) Locked-to-running state
transition for an underdamped Brownian particle in a washboard potential.
The dashed curves define the hysteretic loop of the noiseless case
$\xi(t)\equiv 0$,
with depinning branch starting at
$F_3=\omega_0^2$ and re-pinning branch ending at
$F_1=(4/\pi)\gamma\omega_0$  \cite{borromeo1999b}.
The $T=0+$ step at $F_2$ is represented by a solid curve.
Parameter values: $m = q = 1$ and $\eta = 0.03$.
\label{FigS1}}
\end{figure}

The longitudinal motion of a massive particle on a propagating
substrate $V(x,t)$ can be modelled by replacing
\begin{equation}
\label{S1} V(x) \rightarrow V(x,t)=V(x-vt)
\end{equation}
in the stochastic differential equation (\ref{2.1}). Let us
consider first the sinusoidal wave of Fig. \ref{FigS1}a
\begin{equation}
\label{S2} V(x,t)=-q\cos(x-vt).
\end{equation}
A Galileian transformation, $x(t) \rightarrow y(t)=x(t)-vt$,
allows us to reformulate  Eq. (\ref{2.1}) as
\cite{borromeo1998a,marchesoni2002a}
\begin{equation}
\label{S3} m\ddot y=-\eta \dot y - \eta v - q\sin y + \xi(t).
\end{equation}
Equation (\ref{S3}) describes the Brownian motion in the tilted
washboard potential $V(y)=-q\cos y+\eta v y$, shown in Fig.
\ref{FigS1}(b). This problem was studied in great detail by  Risken
in Ref. \cite{risken1984}. The theme of Brownian motion and
diffusion in periodic potentials has also been widely applied to
describe the transport properties of superionic conductors
\cite{Fulde,Peschel,Geisel}, or for  the evaluation of the thermally
activated escape rates and the corresponding current-voltage
characteristics of damped Josephson junctions \cite{RMP,
StartonovichETAL}. To make contact with Risken's notation, we
introduce the damping constant $\gamma=\eta/m$, the dc driving force
$F=-\gamma v$ and the angular frequency $\omega_0^2=q/m$. The time
evolution of the stochastic process $y(t)$ is characterized by
random switches between a {\it locked} state with zero-mean velocity
and a {\it running} state with asymptotic average velocity $\langle
\dot y \rangle =F/\gamma=-v$. In terms of the mobility
$\mu(T)=\langle \dot y \rangle/F$, locked and running states
correspond to $\gamma \mu=0$ and $\gamma \mu=1$, respectively.
%
In the underdamped, $\gamma \ll \omega_0$, zero-temperature limit,
$T\rightarrow 0$, the stationary dynamics (\ref{S3}) is controlled
by a single threshold $F_2\simeq 3.36 \omega_0\gamma$,
see Fig. \ref{FigS1}c: For $F<F_2$
the particle $y(t)$ sits in one potential well; for $F>F_2$ it
falls down the tilted washboard potential with speed $F/\gamma$;
the $0 \leftrightarrow 1$ jump of $\gamma \mu(T)$ at the threshold
$F_2$ becomes sharper as $T$ tends to zero.

On reverting to Eq. (\ref{S3}) notation, we see immediately that the
thresholds $F_1$-$F_3$ in Fig. \ref{FigS1}(c)
 define three special values of the travelling wave $v$, namely:
\begin{equation}
\label{S4} v_1=\frac{4}{\pi}\sqrt{\frac{q}{m}}, \,\,\,\,\,
v_2\simeq3.36\sqrt{\frac{q}{m}}, \,\,\,\,\, v_3=\frac{q}{\eta}.
\end{equation}
Upon equating $v_1$ and $v_3$ we attain an estimate for the upper
limit $\eta_0$ of the damping constant below which we may expect
to detect a hysteretic cycle, i.e. $\eta_0/m= (\pi/4)\sqrt{q/m}$.
On increasing $\eta$ much larger than $\eta_0$, $v_1$ and $v_2$
merge with $v_3$, which in turn becomes very small
\cite{risken1984}.

The stationary velocity of the Brownian particle $x(t)$ can be
easily determined by inverting the $x \rightarrow y$
transformation, that is
\begin{equation}
\label{S5} v_S\equiv \langle \dot x \rangle =v[1-\gamma \mu(T)]
\equiv b^2(T) v,
\end{equation}
where $v$ is the velocity of the incoming wave and $b(T)$ is the
unknown Stokes' factor. In the presence of noise, no matter how
weak, say $T =0+$, the dynamics of the process is controlled by
$v_2$, only: For $v > v_2$ the process $y(t)$ is in the running
state with $\gamma \mu(T) \simeq 1$, or equivalently the particle
$x(t)$ is subjected to no Stokes' drift, i.e., $v_S\simeq 0$; for
$v < v_2$ the process $y(t)$ is in the locked state with
$\mu(T)\simeq 0$, which corresponds to a dragging speed $v_S\simeq
v$ of the Brownian particle $x(t)$. In the latter case the
particle rides the travelling wave like a surfer ({\it Brownian
surfer} \cite{borromeo1998a}).

The efficiency of the Stokes' drift increases when lowering the
temperature. Moreover, in the low damping regime, $\eta \ll
\eta_0$, it sets on abruptly by tuning the parameters $m$ and $q$
 to appropriate threshold values  -- see Eq. (\ref{S4}).
 Brownian surfers in the overdamped
limit, $\eta \gg \eta_0$, are restricted to either extremely low
frequencies or exceedingly large amplitudes of the travelling
wave, namely $\eta v < q$; for $\eta \rightarrow \infty$
the dragging effect thus becomes less and less efficient.

A massive Brownian particle undergoes Stokes' rectification in the
presence of time and space modulation of its substrate, see also
\cite{borromeo1999a}. For the travelling wave (\ref{S2}) the
displacement of any point  on the substrate averages out to zero,
and so does the spatial average of the substrate deformation at a
given time $t$. However, if we regard $V(x,t)$ as a propagating
elastic wave, the corresponding synchronization of time and space
modulations sustains a net energy transport in the direction of
propagation. In this sense the dynamics (\ref{2.1}) is biased and
Stokes' drift requires no asymmetric profile of the travelling
substrate wave.

For an asymmetric waveform $V(x,t)$, Eqs. (\ref{2.1}) and
(\ref{S1}) describe a travelling {\it ratchet}. Asymmetry makes
then the Stokes' drift problem more intriguing. Suppose we
propagate the piecewise linear potential $V(x)$ of Sec. \ref{SS1}
with constant velocity $v$ according to Eq. (\ref{S1}). Because of
the spatial asymmetry, we can define two threshold speeds
$v_2^{\pm}$ for a wave travelling to the right and to the left,
respectively, with $v_2^-
>v_2^+$. This implies that at low temperatures Stokes' drift to
the right becomes effective e.g. for larger particle masses, viz.
lower substrate amplitudes, than to the left.

However, if the substrate oscillates side-wise with  a constant
speed, with
\begin{equation}
\label{S6} v \rightarrow v(t) = v ~\hbox{sgn}[\cos(\Omega_v t)]
\end{equation}
and $\Omega_v\ll v/l$, the corresponding moving waveform $V(x,t)$
in average transports no energy, but can still induce
rectification because of its asymmetry. Indeed, for $v_2^+ < v <
v_2^-$ the Brownian surfer drifts to the right and the system
works like a ``massive particle sieve''. Note that in the notation of
Eq. (\ref{S3}) such a particle sieve corresponds to a simple
inertial rocked-ratchet \cite{bao2000a,marchesoni1998a}

\section{Quantum Brownian motors}
\label{quantum}

Brownian motors  are typically small physical machines which operate
far from thermal equilibrium by  extracting energy fluctuations,
thereby transporting classical objects on the micro-scale. At
variance with e.g.  biomolecular motors,  certain molecular sized
physical engines necessitate, depending on temperature and the
nature of particles to be transported, a description that accounts
for quantum effects, such as quantum tunnelling and reflection in
the presence of quantum Brownian motion \cite{grabert1988a}. For
this class of quantum Brownian motors recent theoretical studies
\cite{goychuk1998,QBM1,QBM2} have predicted that the transport
becomes distinctly modified as compared to its classical
counterpart. In particular, intrinsic quantum effects such as
tunnelling-induced current reversals \cite{QBM1, linke},
power-law-like quantum diffusion transport laws, and quantum
Brownian heat engines have been observed  with recent, trend-setting
experiments that involve either arrays of asymmetric quantum dots
\cite{linke}, or certain cell-arrays composed of different Josephson
junctions \cite{majer}.

\begin{figure}[htb]
\includegraphics[width=.70\textwidth]{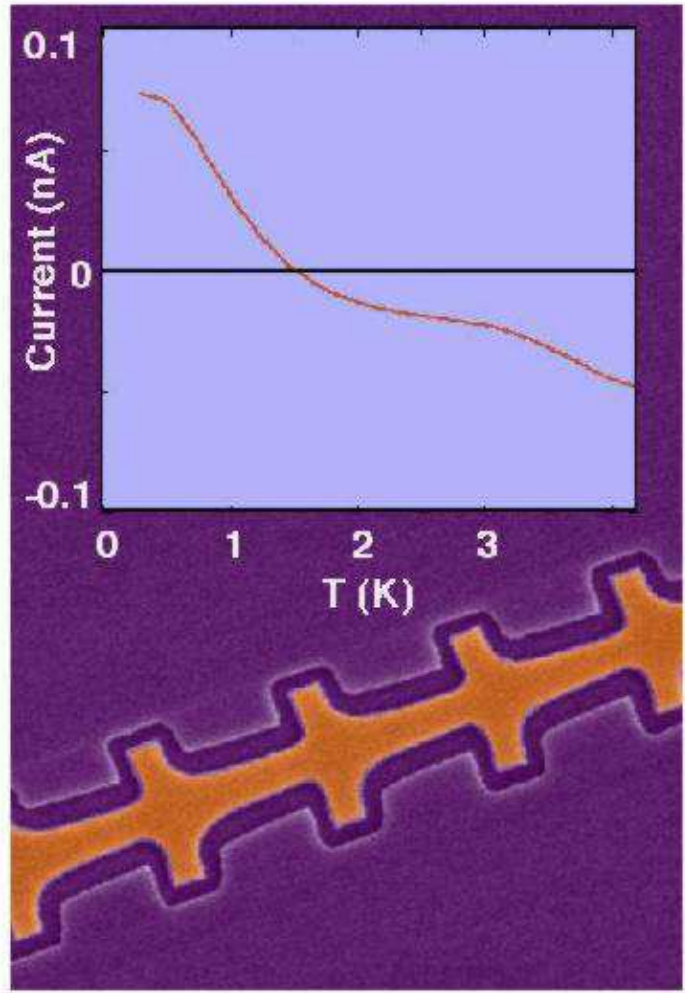}
\centering \hfil
\caption{In a {\it quantum Brownian motor}, being
driven by an adiabatically varying ac-rocking voltage, quantum
tunnelling can contribute to the electron current. Due to the
underlying asymmetric potential structure, the two components to the
time-averaged net current are of opposite sign \cite{QBM1}. The
strength of the two contributions can be tuned individually by
temperature. This causes a tunnelling-induced current reversal
(occurring near $1.5$K in the top graph) \cite{linke} that can be
exploited to direct electrons along {\em a priori} designed routes.
Below the measurement graph is a scanning  electron micrograph of
the used quantum ratchet device. Figure provided by Heiner Linke,
University of Oregon.} \label{fig:4}
\end{figure}

In contrast to the {\em classical} description, the theory  for {\em
quantum} Brownian motors (as well as corresponding experiments) is
much more demanding. This is mainly rooted in the fact that one has
to master the mutual interplay of (i) quantum mechanics, (ii)
quantum dissipation, and (iii) non-equilibrium driving. Any of these
three aspects alone is already not straightforward to accommodate
theoretically. In particular, the theoretical description of
non-equilibrium, dissipative quantum Brownian motors schemes is
plagued by difficulties such as: (a) the commutator structure of
quantum mechanics occurring in the Hilbert space of the combined
system plus the bath(s), (b) the  description of quantum dissipation
that at all times necessitates consistency with the Heisenberg
relation and the entanglement features between system and
environment(s), (c) the correct treatment of quantum detailed
balance \cite{talkner1986a} in equilibrium,  so that no quantum
Maxwell demon is left alive when all applied non-equilibrium sources
are ``switched-off'', to name only a few of the main causes of
possible theory-related pitfalls.

The present state of the art of the theory is thereby characterized
by specific restrictions such as, e.g., an adiabatic driving regime,
a tight-binding description, a semiclassical analysis, or
combinations thereof \cite {QBM1,QBM2}. As such, the study of
quantum Brownian motors is far from being complete and there is
plenty of room and an urgent need for further developments. A
particular challenge for theory and experiments are quantum Brownian
motors that are built from bottom up on the nano-scale. First
results for quantum Brownian rectifiers based on infrared irradiated
molecular wires have recently been investigated in
\cite{lehmann2002a}. In those quantum systems one employs coherent,
driven tunnelling \cite{Grifoni1998} through tailored asymmetric
nano-structures, in combination with dissipation due the coupling to
macroscopic fermionic leads, which are kept at thermal equilibrium.

\begin{figure}[htb]
\includegraphics[width=.45\textwidth]{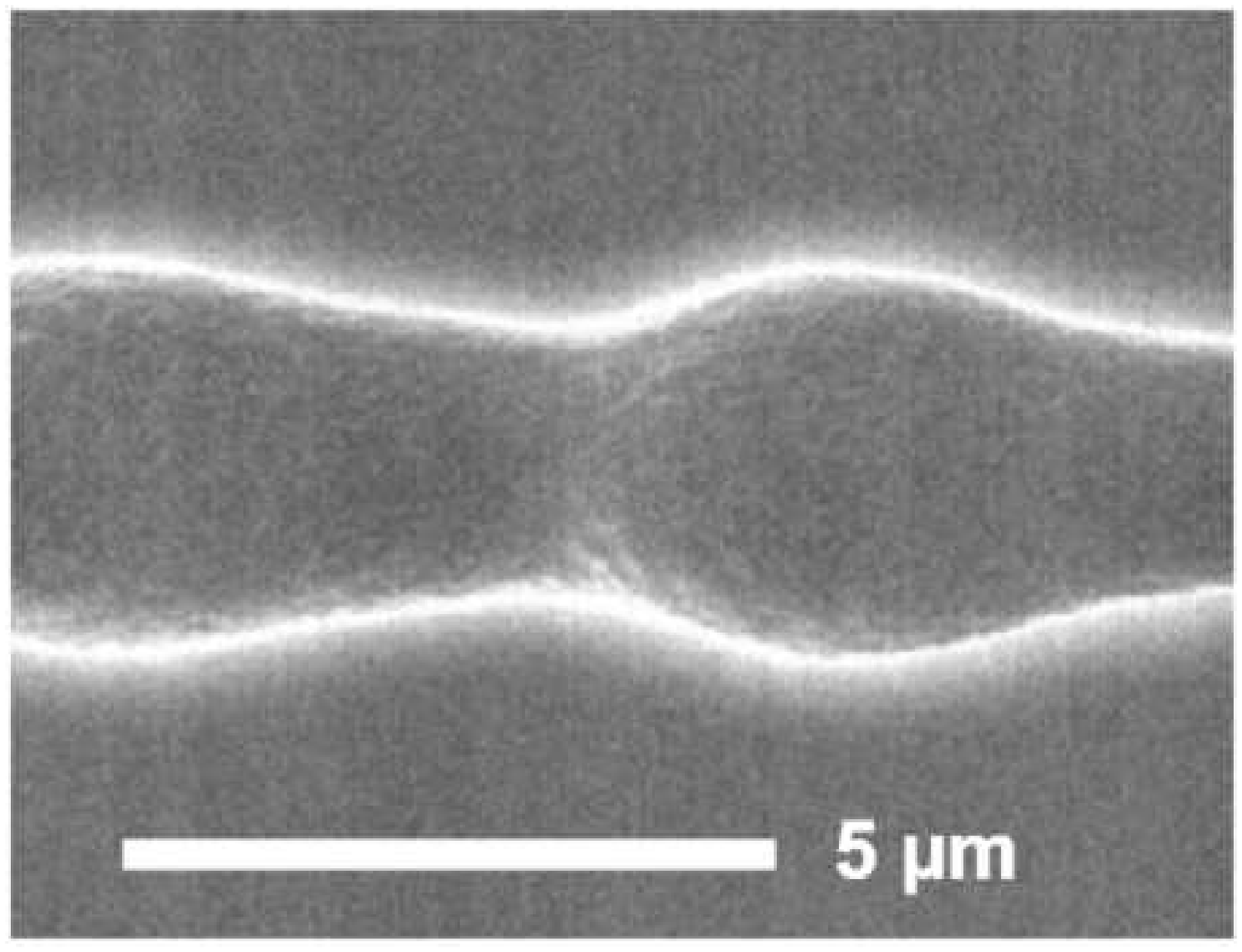}~a)
\hfil
\includegraphics[width=.45\textwidth]{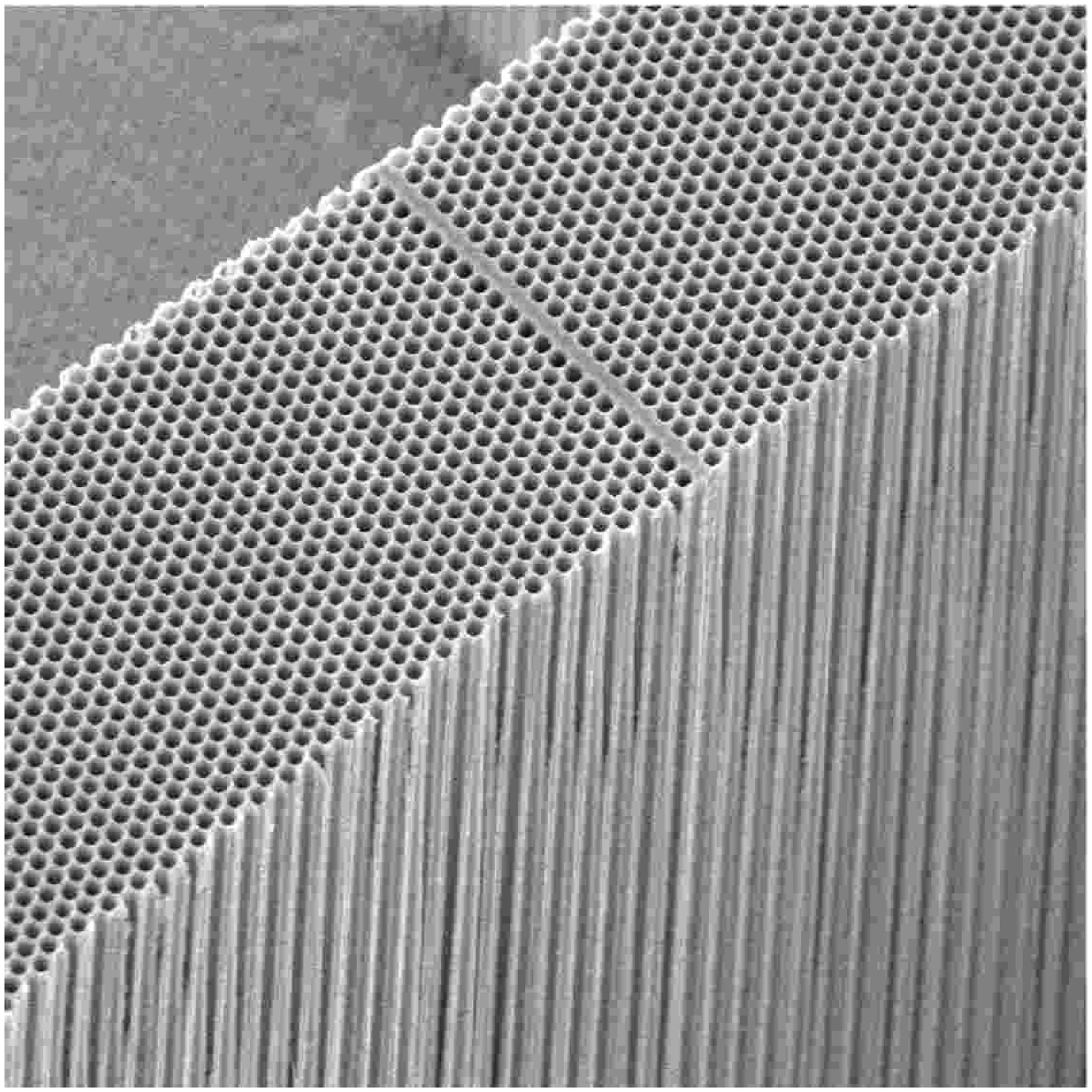}~b)
\caption{Panel (a): A scanning-electron-microscope picture of a
single pore  with a ratchet-shaped  (i.e., asymmetric) periodic
variation  of the diameter along its axis; the length of one period
is  $8.4 \mu \text{m}$. Panel (b): Scanning-electron-microscope
picture of a silicon wafer which is pierced by a huge number of
practically identical pores with pore distances of 1.5 mm and pore
diameters of 1 mm. This illustrates the enormous potential for
separation with a parallel three-dimensional ratchet-architecture.}
\label{fig:5}
\end{figure}

As a typical example where theory and experiment have met, we
discuss the case of a rocking quantum ratchet as depicted in the
bottom panel in Fig. \ref{fig:4}. It is  known that for a slowly
rocked classical  Brownian motor (adiabatic regime)
\cite{bartussek1994a, magnasco1993a}, the noise-induced transport
does not exhibit a reversal of current direction. Such a reversal
occurs only in the non-adiabatic rocking regime at  higher driving
frequencies \cite{bartussek1994a}. This  very situation changes
drastically when quantum tunnelling enters into the dynamics. A true
benchmark for a quantum behavior of an adiabatically rocked Brownian
motor is then the occurrence of a tunnelling-induced reversal at low
temperatures, as theoretically predicted in \cite{QBM1}. This
characteristic feature has been  experimentally verified with an
electron quantum rocking Brownian motor composed of a
two-dimensional gas of electrons moving within a fabricated,
ratchet-tailored hetero-structure of a GaAs/AlGaAs interface
\cite{linke}, see the top panel of Fig. \ref{fig:4}. This current
reversal indicated the existence of parameter configurations where
the quantum Brownian motor current vanishes. In the neighborhood of
these system configurations we consequently can devise a quantum
refrigerator that separates ``cold'' from ``hot'' electrons in
absence of currents \cite{linke}.

\section{Recent applications}
\label{new} Over the last decade or so, many theoretical schemes
and experimental implementations of Brownian motors have been
devised \cite{BM,BMRH}. Several recent applications use an external
rocking force, of electric or mechanical origin, as a tunable
control.

\begin{figure}[htb]
\includegraphics[width=.45\textwidth]{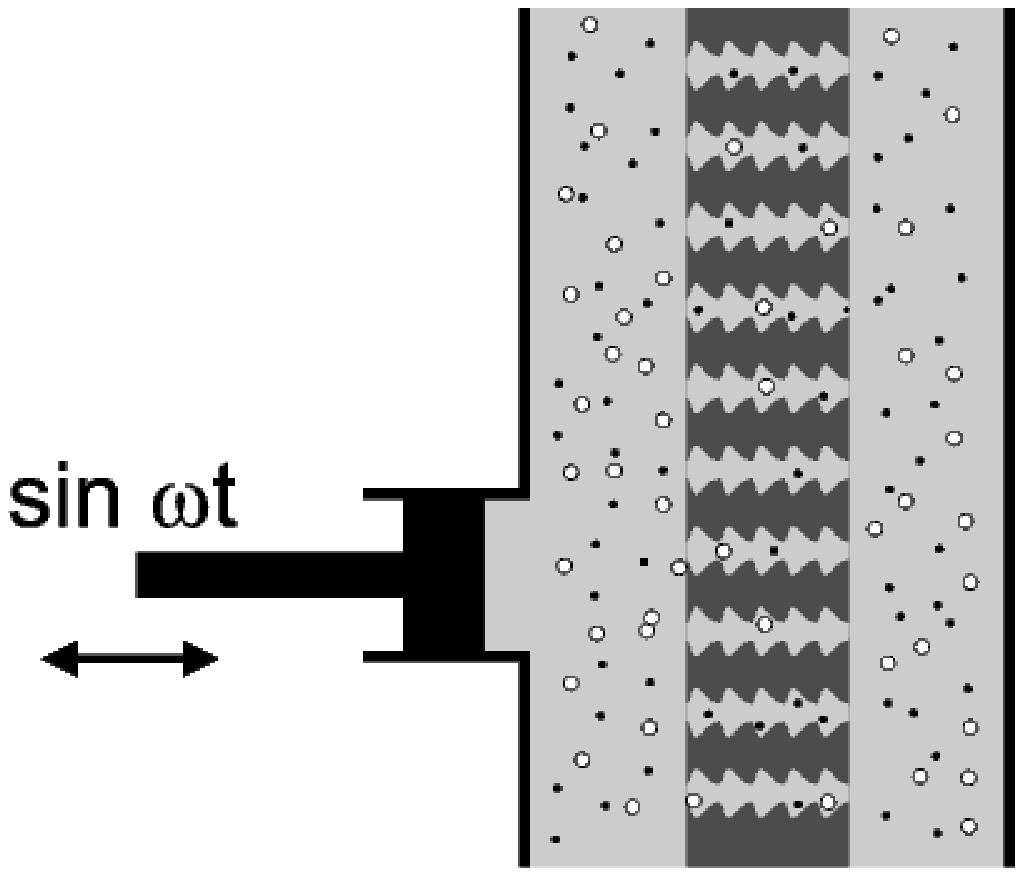}~a)
\hfil
\includegraphics[width=.50\textwidth]{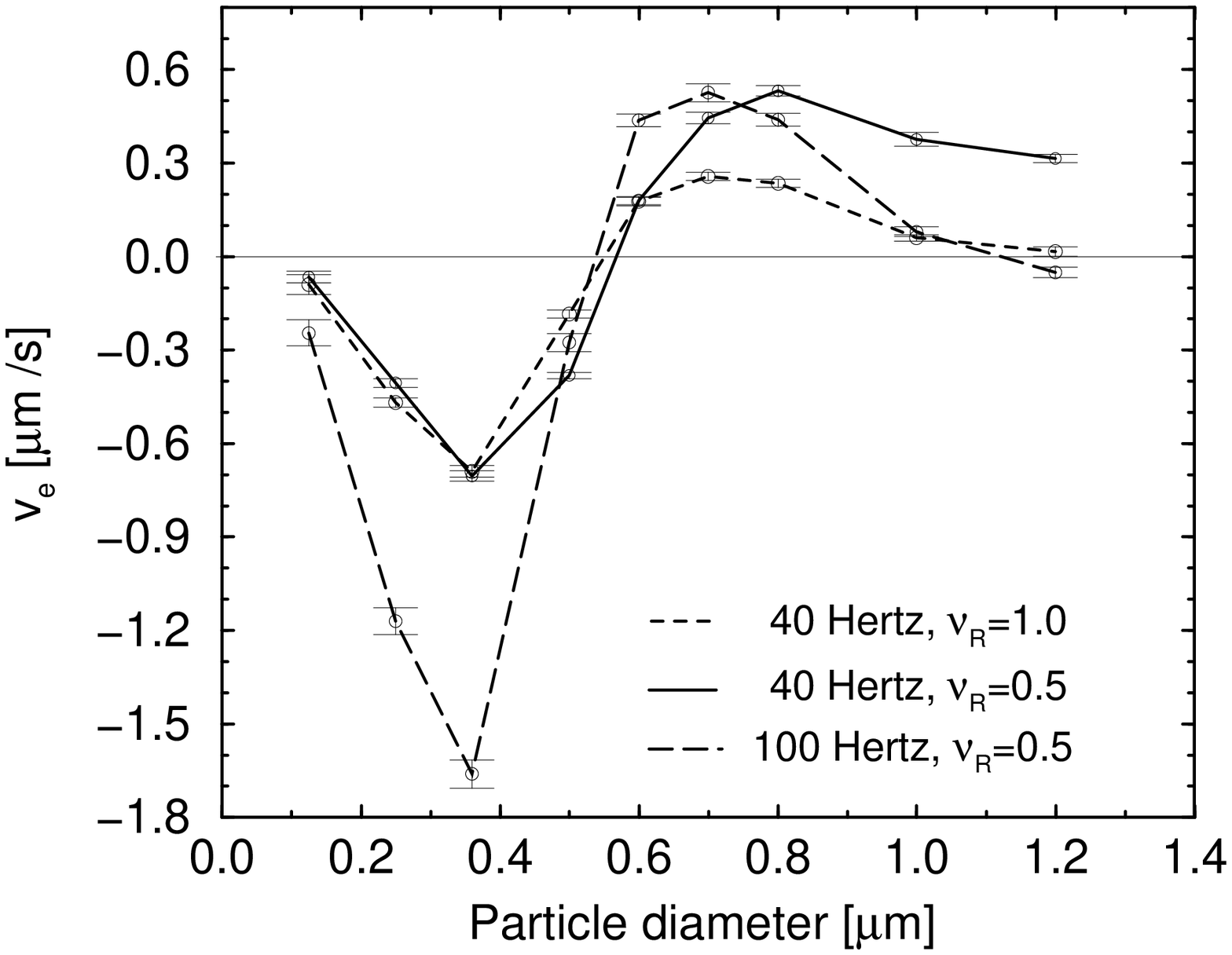}~b)
\caption{Panel (a): Concept of a microfluidic drift ratchet.
Schematic cross section through the
 plane of the experimental setup. A macro-porous silicon
wafer is connected at both ends to basins. The pores with their
ratchet-shaped profile are schematically indicated in dark grey. The
basins and the pores are filled with liquid; micrometer-sized
particles of two different species are indicated. The fluid is
pumped back and forth by a pumping device, indicated by the piston
on the left hand side. Figures provided by Christiane Kettner {\it
et al.} \cite{kettner2000a}. Panel (b): Average Brownian motor
induced particle current $v_e$  versus  particle diameter for
various driving frequencies $\omega/2\pi$  and  viscosities
(relative to water) $\nu_R$. Particularly note the very sharp
velocity reversal around $ 0.5 \,\mu$m. For further details, see
Kettner {\it et al.} \cite{kettner2000a}}.
 \label{fig:6}
\end{figure}

\begin{figure}[htb]
\hfil
\includegraphics[width=.95\textwidth]{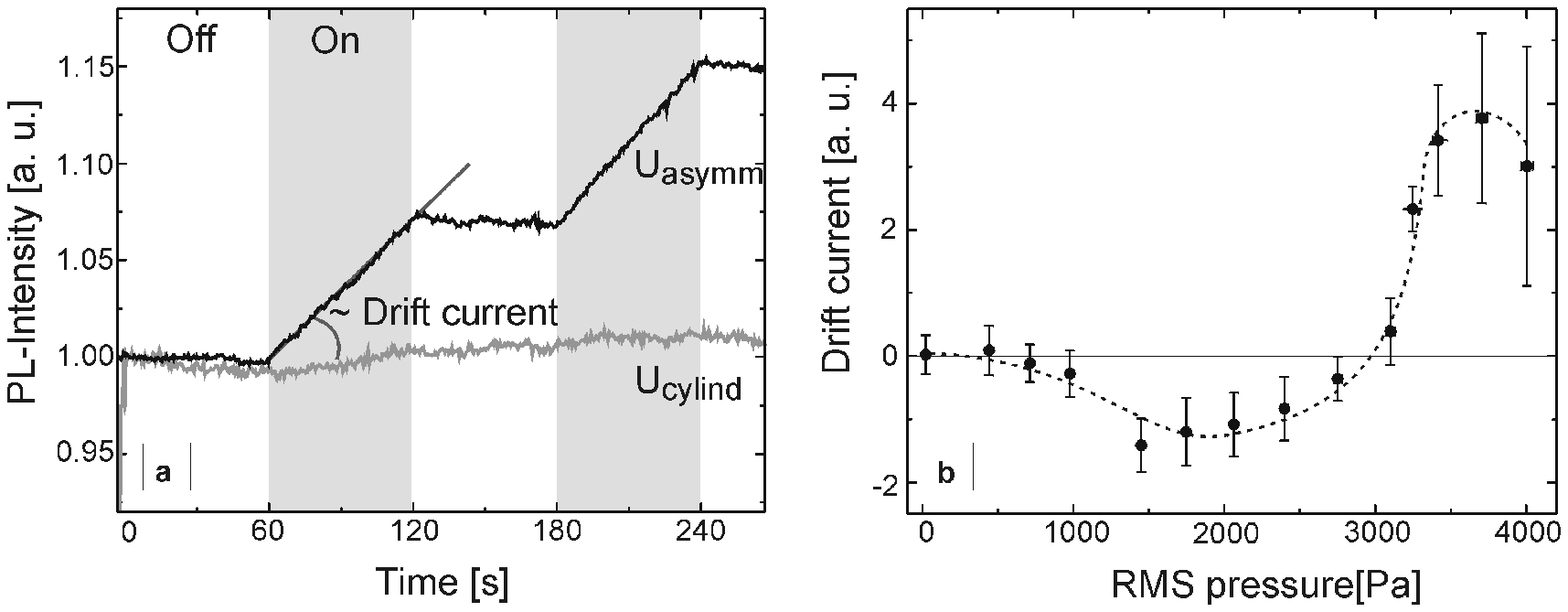}
\caption{Parallel acting Brownian motors: Asymmetric pores  in a
macroporous silicon membrane containing ca. 1.7 million pores act as
massively parallel Brownian motors, cf. in Ref. \cite{muller2003a}.
When the pressure oscillations of the water are switched on, the
photoluminescence signal and thus the number of particles in the
basin located to the right, see in panel (a) of Fig. \ref{fig:6}
increases linearly. For symmetrically, cylindrical-shaped pores no
systematic drift is observed, see in a). The net transport behavior
is strongly dependant on the applied pressure amplitude and shows
qualitatively the theoretically predicted current inversion b). The
pressure oscillations are toggled on and off each 60 s. The
experimental parameters used are as follows: the suspended
luminescent polystyrene spheres  in water possess a diameter of 0.32
$\mu \text{m}$, the pressure oscillation frequency is 40 Hz and the
applied root mean square (r.m.s.) pressure during the `on' phase of
2000 Pa. The number of etched modulations in a single pore was 17.
[Image: Max-Planck-Institute of Microstructure Physics]}
\label{fig:7}
\end{figure}

A few fascinating examples are the light powered single-molecule
opto-mechanical cycle, experimentally studied by H.~E. Gaub and
collaborators \cite{gaub2002a}, the use of colloidal suspensions of
ferromagnetic nanoparticles \cite{engel2003a},  ratchet devices that
control the motion of magnetic flux quanta in superconductors
\cite{cooperative2,cooperative3,natmat2003,villegas2003a}, or the
Brownian motor induced clustering of a vibrofluidized granular gas
yielding the phenomenon of a {\it granular fountain} and the {\it
granular ratchet} transport perpendicular to the direction of
unbiased energy input \cite{lohse2004a}. Yet another (Brownian
motor)-related phenomenon is the emergence of paradoxical motion of
noninteracting, driven Brownian particles exhibiting an {\it
absolute negative mobility} and corresponding current reversals
\cite{eichhorn2002a}. Note that an absolute negative mobility
implies that the response is {\em opposite} to the applied force
that is applied around the origin of zero force; as such,  this
phenomenon must be distinguished from so-called differential
negative mobility which is typified by a negative-valued slope of
the response-force characteristics {\it away} from  the origin.

Here, we discuss the prominent potential of a microfluidic
realization of Brownian motors that  can be used to separate
particles with large separation power and in short times. The set-up
of the device is depicted in Figs. \ref{fig:5},  \ref{fig:6}. It
consists of a three-dimensional array of asymmetric pores, see Fig.
\ref{fig:5}a), b) in which a fluid such as water containing some
immersed, suspended polystyrene particles is pumped back and forth
with no net bias (!), see Fig. \ref{fig:6}a). Due to the asymmetry
of the pores, the fluid develops, however,  asymmetric flow patterns
\cite{kettner2000a}, thus providing the ratchet field of force in
which a Brownian particle of {\it finite size}  can both, (i)
undergo Einstein diffusion into liquid layers of differing speed,
and/or (ii) become reflected asymmetrically from the pore walls.
Both mechanisms will then result in a driven non-equilibrium net
flow of particles.

The numerical evaluation of the Brownian motor current then yields a
rich behavior, featuring an amazingly steep current reversal as a
function of the particle size,  see Fig. \ref{fig:6}b). Note
 that the direction of the net flow cannot be easily guessed
{a priori}; indeed, the direction of the Brownian motor current is
determined by the interplay of the Navier-Stokes flows in this
tailored geometry and hydrodynamic thermal fluctuations. This
proposal for a microfluidic ratchet-based pumping device has
recently been put to work successfully with experiments
\cite{muller2003a}: the experimental findings are in good
qualitative agrement with theory; but more work is required to
achieve detailed quantitative agreement.

Remarkably, this device has advantageous three-dimensional scaling
properties \cite{kettner2000a,muller2000a}: a massively parallel
architecture composed of ca. 1.7 million  pores, cf. Fig.
\ref{fig:5}b), is capable to direct and separate micron sized
suspended objects very efficiently, see in Fig. \ref{fig:7}. These
type of devices have clear potential for bio-medical separation
applications and therapy use.
\begin{figure}[htb]
\includegraphics[width=.70\textwidth]{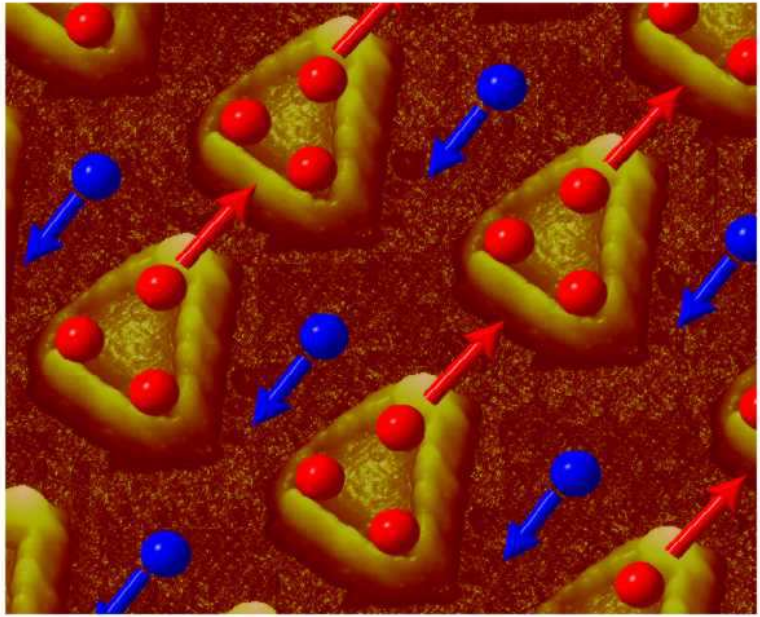}
\centering \hfil
\caption{Superconducting Niobium film grown on an array of Nickel
triangles. The magnetic flux quanta, or vortices, shown as balls,
can be separated in two groups: (i) pinned vortices, shown in red,
which move from one triangular-shaped pinning trap to another one
and, thus, these are directly affected by the pinning potential; and
(ii) interstitial vortices, shown in blue, which move in-between
triangles, and do not directly interact with the pinning traps.
However, the interstitial vortices can indirectly feel the spatial
asymmetry via their interactions with the pinned vortices. This
problem can be mapped into the similar system of two species of
repulsive particles in which one type or species of particles
directly interacts with the spatially-asymmetric substrate. The
other type of particle (interstitial vortices here, shown in blue)
is insensitive to the substrate, at least in a direct manner. It has
been shown in that those particles (assigned red) subject to the
substrate potential create an effective asymmetric potential, with
the opposite asymmetry or opposite polarity, for the other (blue)
particles. When all the particles are subjected to an ac drive
force, this ``inverted-polarity'' potential rectifies the motion of
the blue particles (interstitial vortices in our case) in one
direction, and the original pinning potential rectifies the motion
of the red particles (pinned vortices) along the opposite direction,
because the latter feel a potential with opposite polarity. Figure
provided by Jose Vicent, Universidad Complutense de Madrid.}
\label{fig:8}
\end{figure}

Another area of growth regarding applications of Brownian motors to
micro-devices involve the control of the motion of quantized flux
quanta in superconductors
\cite{cooperative2,cooperative3,natmat2003,villegas2003a}. For
instance, the authors of Ref.~\cite{villegas2003a} fabricated a
device that controls the motion of flux quanta in a Niobium
superconducting film grown on an array of nanoscale triangular
pinning potentials, cf. Fig. \ref{fig:8}.

The controllable rectification of the vortex motion is due to the
asymmetry of the fabricated magnetic pinning centers. The reversal
in the direction of the vortex flow is explained theoretically by
the interaction between the vortices trapped on the magnetic
nanostructures and the interstitial vortices. The applied magnetic
field and input current strength can tune both the polarity and
magnitude of the rectified vortex flow. That ratchet system is
explained and modeled theoretically considering the interactions
between particles. This device allows a versatile control of the
motion of vortices in superconducting films. Simple modifications
and extensions of it
\cite{cooperative2,cooperative3,natmat2003,villegas2003a}would
allow the pile-up (magnetic lensing), shaping, or "sculpting" of
micromagnetic profiles inside superconductors. Vortex lenses made
of oppositely oriented asymmetric traps would provide a strong
local increase of the vortex density at its focus regions.
Extensions of these types of systems
\cite{cooperative2,cooperative3,natmat2003,villegas2003a} could
allow the motion control of interacting particles in colloidal
suspensions, and interacting particles in micro-pores, and not
just controlling the motion of flux quanta. These systems provide
a step toward the ultimate control of particle motion in tiny
microscopic devices.

\section{Conclusion}
With this work we commemorate some intriguing features of the rich
physics of Brownian motion which Albert Einstein pioneered 100 years
ago. We can assess that the physics of classical and quantum
Brownian motion and its use for technological are still very much
under investigation. One main lesson to be learned from Einstein's
work is that rather than fighting thermal Brownian motion we should
put it to constructive use: Brownian motors take advantage of this
ceaseless noise source to efficiently direct, separate, pump and
shuttle particles reliably and effectively.

\begin{acknowledgement}
This work has been  supported by the Deutsche
Forschungsgemeinschaft, SFB-486, project A-10 (PH);
DFG-Sachbeihilfe HA 1517/13-4 (PH); and the
Baden-W\"urttemberg-Bayern initiative on quantum information
processing (PH). FN acknowledges support from the NSA and ARDA
under AFOSR contract No. F49620-02-1-0334 and the NSF grant
EIA-0130383.
\end{acknowledgement}

\end{document}